\documentclass[a4paper,11pt]{article}
\usepackage{pos}
\usepackage{hyperref}

\newcommand{\slz}{\text{SL}(2,\mathbb{Z})}

\newcommand{\mc}{\mathcal}

\newcommand{\sbar}{\bar{S}}

\renewcommand{\epsilon}{\varepsilon}

\title{On (de Sitter) Vacua
in Heterotic String Theory}

\author*[a]{Nicole Righi}

\affiliation[a]{Physics Department, King's College London,\\ Strand, London, WC2R 2LS, U.K.}

\emailAdd{nicole.righi@kcl.ac.uk}

\abstract{Heterotic toroidal orbifolds are a powerful framework for testing our understanding of string vacua. We review recent advances in heterotic model building, from constructing new vacua to proving novel de Sitter no-go theorems. We also illustrate a potential loophole that could pave the way for the realisation of de Sitter vacua within heterotic string theories. This involves the introduction of non-perturbative, stringy, dilaton-dependent corrections to the K\"ahler potential. We discuss how recently found non-BPS heterotic instantons might provide the necessary corrections.}

\FullConference{2nd Training School and General Meeting of the COST Action COSMIC WISPers  (CA21106) (COSMICWISPers2024)\\
 10-14 June 2024 and 3-6 September 2024\\
Ljubljana (Slovenia) and Istanbul (Turkey)\\}

\begin{document}
\maketitle

\section{The positive $\Lambda$}
The accelerating expansion of the universe~\cite{SupernovaSearchTeam:1998fmf} and the nature of the dark energy driving it remain a great challenge for theoretical physics. The simplest model that aligns with observations is that our universe is in a de Sitter (dS) phase, and so dark energy corresponds to a cosmological constant in Einstein’s equations. One might hope that natural explanations can be found in a UV-complete theory such as string theory. However, the difficulty of weakly coupled string theory in providing the origin and scale of the cosmological constant has recently sparked speculations about alternative models to address the apparent lack of (fully controlled) dS vacua in string theory. These models mostly rely on quintessence scenarios of scalar fields slowly rolling towards a runaway at infinity, the generic and naive outcome of string theory compactifications. However, these models not only require more fine-tuning compared to dS models to explain our universes~\cite{Cicoli:2018kdo} but also lack compelling experimental evidence~\cite{DESI:2024jis}. 

Constructing a vacuum from string theory involves a complex interplay of perturbative and non-perturbative quantum corrections to stabilise the moduli. Some of these corrections are well-studied, while others are difficult to tackle at present or completely unknown, hence the claimed lack of full control on known vacua. To finally claim control, it is of great importance to keep studying quantum corrections both purely from the 4D EFT point of view, as well as in the full 10D picture of string theory to completely understand their origin. Here, we show from a bottom-up perspective how the inclusion of new, non-perturbative effects can dramatically change the structure of vacua in heterotic string theories. At the end, we provide a possible 10D explanation of these effects and comment on prospects.

\section{Moduli stabilisation in toroidal orbifolds}
Heterotic string theories are a remarkable playground where we can test our understanding of string theory and quantum gravity and their relation with our world. At the technical level, we have powerful tools such as the 4D $\mathcal{N}=2$ duality with type II theories and the $\mathcal{N}=1$ with type I theory, modular symmetries once we compactify on tori, and higher-supersymmetric subsectors and worldsheet CFT computation of quantum corrections which give great control over the EFT without the need of a large volume/weak coupling regime. Over the years, there has been outstanding work in the literature on Standard Model and vacua constructions, and of interest for this note are the many no-go theorems for de Sitter vacua~\cite{Green:2011cn,Gautason:2012tb,Kutasov:2015eba,Quigley:2015jia,Leedom:2022zdm}.

For these reasons, we revisit moduli stabilisation and (de Sitter) vacua in heterotic string theories on toroidal orbifolds, providing additional no-go theorems and a way to evade them~\cite{Leedom:2022zdm}. Moduli stabilisation can be carried out on toroidal orbifold compactification due to the reduced amount of supersymmetry in 10D. The flux structure of heterotic theories greatly differs from those of type~ II, which makes flux compactification on Calabi-Yau threefolds difficult to treat and insufficient at fully stabilising the moduli. Instead, one can rely on the $\slz$ symmetries carried by tori and stabilise (at least some) moduli at the fixed points, i.e. those points fixed by cyclic subgroups of $\slz$. For clarity, we focus on a simple setup: we consider a 6D orbifolded torus parametrised by an overall K\"ahler modulus $T=a+it$, and neglect complex structure moduli 
and matter fields. Most importantly, we will be concerned with stabilising the dilaton supermultiplet $S=1/g_s^2 + i \theta$, whose value determines the string coupling constant $g_s$. 
This model must be viewed more as a proof of concept than a concrete proposal for describing our universe, an important task that would require further model building. 

The modular symmetry of the torus dictates most of the rules to build the EFT, starting with the transformation of the K\"ahler modulus, $  T\rightarrow 
    \left(aT+b\right)\left(cT+d\right)^{-1},
    $
for which the fixed points are $T=i$ and $T= e^{2\pi i/3}\equiv \rho$. We want to formulate appropriate superpotential $W$ and K\"ahler potential $K$ to compute the F-term supergravity potential 
\begin{equation}
     V = 
     e^{K} \left(K^{i\bar j} D_iW  D_{\bar j}\bar W - 3W\bar W\right)=e^{\mc{G}} \left(\mc{G}_i\mc{G}^{i\bar j} \mc{G}_{\bar{j}} - 3\right)\,,
\label{eq:sugrapot}    
\end{equation}
where $D_i W=\partial_i W + W \partial_i K$, $i=T,\,S$, and we introduced the modular invariant supergravity function $\mathcal{G} = K + \ln\lvert W\rvert^2$. The K\"ahler potential is parametrised as
\begin{equation} 
    K= k -3\ln\left(-i(T-\bar T)\right)\,\,\,\, \text{ where } \,\,\,\,k\equiv k(S,\bar{S}) = -\ln(S+\bar{S}) + \delta k_{np}(S,\bar{S})\,.
    \label{eq:kahpot1}
\end{equation}
Under modular transformations 
(assuming that $k$ is inert under $\slz$), 
the K\"{a}hler potential undergoes a K\"{a}hler transformation $
K\rightarrow K + 3\ln(cT+d)+3\ln(c\bar{T}+d)$: it is a weight $(3,3)$ modular form. This restricts the form the superpotential can take to keep $\mathcal{G}$ modular invariant, namely $W$ must be a weight $(-3,0)$ modular form. The most general nonperturbative superpotential satisfying such requirement has the form~\cite{Cvetic:1991qm}
\begin{equation}
 W(S,T) = \frac{\Omega(S)H(T)}{\eta^6(T)}\,,
 \label{eq:superpot1}
\end{equation}
where $\Omega(S)\sim e^{-S}$ arises from gaugino condensation of some subgroups of the 10D gauge sector, $\eta(T)$ is the Dedekind eta and $H(T)$ is a modular function with a potentially non-trivial multiplier system; if regular in the fundamental domain, $H(T)$ can be parametrised in the most general way as~\cite{10.2307/1968796}
\begin{equation}
    H(T) = \bigg(\frac{G_4(T)}{\eta^8(T)}\bigg)^n\bigg(\frac{G_6(T)}{\eta^{12}(T)}\bigg)^m \mathcal{P}(j(T))\,,
   \label{eq:Hpara}
\end{equation}
with $G_n(T)$ the weight $(n,0)$ holomorphic Eisenstein series, $j(T)$ the $j$-invariant and $\mathcal{P}(x)$ is a polynomial. Using (\ref{eq:kahpot1}) and (\ref{eq:superpot1}) one can compute (\ref{eq:sugrapot}) and express it in terms of the function \begin{equation}
A(S,\bar{S}) = \frac{k^{S\sbar}F_S\bar{F}_{\sbar}}{\lvert W\rvert^2} =\frac{k^{S\sbar}\lvert\Omega_S + K_S\Omega\rvert^2}{\lvert\Omega\rvert^2}\,.\end{equation}
Note that this function parametrises the breaking of supersymmetry in the dilation sector through non-trivial dilaton F-terms $F_S\neq 0$. 

When $\delta k_{np}(S,\bar{S})$, $A(S,\bar{S})=0$ and the theorems proven in~\cite{Leedom:2022zdm} ensure that the potential will never have a dS minimum;  hence one can find a loophole by requiring $A(S,\bar{S})>0$, which seems to be almost the only loophole left to realise de Sitter vacua in heterotic toroidal orbifold compactifications. Different conditions to stabilise the K\"ahler moduli sector, which correspond to different values of $(m,\,n)$ in (\ref{eq:Hpara}), translate into bounds on the values $A(S,\bar{S})$ must take to have a metastable dS minimum. These differ depending on whether the K\"ahler modulus $T$ is stabilised at the fixed points or on the rest of the fundamental domain of $\slz$.

A nontrivial $A(S,\bar{S})$ is realised when $\delta k_{np}(S,\bar{S})\neq 0$. How should we parametrise $\delta k_{np}(S,\bar{S})$ explicitly?  Can a $\delta k_{np}(S,\bar{S})$
of the right magnitude exist? Following~\cite{Silverstein:1996xp}, ``K\"ahler Stabilized Models" of heterotic particle phenomenology introduce a term $\delta k_{np}(S,\bar{S})\sim e^{-1/g_s}$ to stabilise $S$ in a Minkowski vacuum and avoid a runaway in the string coupling~\cite{Gaillard:2007jr}. These effects dominate over non-perturbative gauge theory ones, which scale as $e^{-1/g_s^2}$. However, $e^{-1/g_s}$ stringy effects are poorly understood and deserve thorough investigation. Indeed, their impact on heterotic vacua in the effective 4D theories considered above is remarkable and, if consistent, could provide a way to generate dS vacua in heterotic string theories via truly stringy features~\cite{Leedom:2022zdm}. 

Finally, we note that the value of the potential at the minimum is exponentially sensitive to the string coupling. Achieving scale separation and, eventually, a small cosmological constant might be in tension with the magnitude of $A(S,\bar{S})$ needed for the uplift described above. We will get back to this intriguing investigation once the exact form of the function $A(S,\bar{S})$ is determined.



\section{Non-perturbative effects and 10D instantons}
Non-perturbative effects falling off like $e^{-1/g_s}$ are inherently stringy and typically arise from D-branes in theories with open strings. Considering them in heterotic models might seem surprising, given that these theories lack D-branes. Nonetheless, S.~Shenker showed that \emph{every} closed string theory must also incorporate open strings solely from the scaling properties of closed string amplitudes~\cite{Shenker:1990}. Then, ref.~\cite{Green:2016tfs} demonstrated that corrections proportional to $e^{-1/g_s}$ are present in the $R^4$-term of the 9D and 10D actions, the latter at least for Spin$(32)/\mathbb{Z}_2$ theories. In particular, they show up resummed in the real analytic Eisenstein series $E_{3/2}(i g_s^{-1})$. Recently, the 10D origin of these effects was found in non-BPS stringy instantons, non-perturbative configurations whose endpoints consistency can be described by mixing worldsheet and spacetime degrees of freedom~\cite{Alvarez-Garcia:2024vnr}, expanding on the construction of~\cite{Polchinski:2005bg}. 

In the heterotic Spin$(32)/\mathbb{Z}_2$ theory, the existence of these instantons is signalled by the non-trivial homotopy group $\pi_9($Spin$(32)/\mathbb{Z}_2)=\mathbb{Z}_2$, 
which denotes the presence of a gauge profile (i.e. spacetime gauginos) supported on spacetime filling NS9-branes and means that \emph{pairs} of instantons contribute to the action even if the whole tower is unstable. 
The heterotic $($E$(8)\times$E$(8))\rtimes\mathbb{Z}_2$ theory
 instead does not carry a gauge bundle 
 as its homotopy class is trivial. However, $e^{-1/g_s}$ terms can originate from unstable gravitational configurations and would involve a mixing of worldsheet fermions and 10D gauginos, dilatino, and gravitino on a non-trivial spacetime.

Therefore, upon compactification, $e^{-1/g_s}$ corrections in 4D can potentially originate from D-strings~\cite{Hull:1997kt,Hull:1998he}, $0$-branes~\cite{Polchinski:2005bg,Kaidi:2023tqo}, and from the ($-1$)-branes~\cite{Alvarez-Garcia:2024vnr} which are already present in 10D and explain Shenker's argument for heterotic string theories. While so far we have only pointed out the objects responsible for $e^{-1/g_s}$ corrections and described their nature, further work is required to properly establish their actual contribution to the 4D EFT. However, let us remark that such corrections would affect non-holomorphic quantities such as the K\"ahler potential, justifying the bottom-up assumption in \eqref{eq:kahpot1}. 

Overall, these results show the importance of digging further into all the possible quantum corrections emerging from string theory and their implications for understanding the cosmological constant, in particular by venturing inside the moduli space and beyond BPS objects, since we just saw that non-BPS quantities do contribute to the action.

\acknowledgments
\noindent I thank Rafael \'Alvarez-Garc\'ia, Christian Knei{\ss}l, Alexander Westphal and in particular Jacob M. Leedom for collaborations on the topics presented in this publication, based upon work from COST Action COSMIC WISPers CA21106, supported by COST (European Cooperation in Science and Technology).
My work is supported by a Leverhulme Trust Research Project Grant RPG-2021-423.

\bibliographystyle{JHEP}
\bibliography{ref}

\end{document}